\documentclass[12pt]{article}
\usepackage{graphicx}
\usepackage{subfig}
\usepackage{lineno}
\usepackage{amsmath}
\usepackage{multirow,bigdelim}
\usepackage{booktabs}
\usepackage{placeins}

\newcommand\pubnumber{ATL-PHYS-PROC-2019-015}
\newcommand\pubdate{\today}

\def\institute{Physikalisches Institut\\
Universit\"at Bonn, 53115 Bonn, Germany}

\def\support{\footnote{Supported by European Research Council grant
ERC-CoG-617185.}}

\makeatletter
\def\blfootnote{\xdef\@thefnmark{}\@footnotetext}
\makeatother

\def\copyright{\blfootnote{Copyright 2019 CERN for the benefit of the ATLAS Collaboration. CC-BY-4.0 license.}}

\def\Title#1{\begin{center} {\Large #1 } \end{center}}
\def\Author#1{\begin{center}{ \sc #1} \end{center}}
\def\Address#1{\begin{center}{ \it #1} \end{center}}

\newcommand\pubblock{\rightline{\begin{tabular}{l} \pubnumber\\
         \pubdate  \end{tabular}}}
\newenvironment{Abstract}{\begin{quotation}  }{\end{quotation}}
\newenvironment{Presented}{\begin{quotation} \begin{center} 
             PRESENTED AT\end{center}\bigskip 
      \begin{center}\begin{large}}{\end{large}\end{center} \end{quotation}}
\def\Acknowledgements{\bigskip  \bigskip \begin{center} \begin{large}
             \bf ACKNOWLEDGEMENTS \end{large}\end{center}}

\def\beq{\begin{equation}}
\def\eeq#1{\label{#1}\end{equation}}
\def\eeqn{\end{equation}}

\def\beqa{\begin{eqnarray}}
\def\eeqa#1{\label{#1}\end{eqnarray}}
\def\eeqan{\end{eqnarray}}

\let\bar=\overbar

\def\Dslash{\not{\hbox{\kern-4pt $D$}}}
\def\dslash{\not{\hbox{\kern-2pt $\del$}}}

\def\msb{{\bar{\ssstyle M \kern -1pt S}}}

\def\ttZ{\ensuremath{t \bar{t} Z}}

\def\ttW{\ensuremath{t \bar{t} W}}
\def\GeV{\ensuremath{\text{Ge\kern -0.1em V}}}
\def\TeV{\ensuremath{\text{Te\kern -0.1em V}}}

\begin{document}
\begin{titlepage}
\pubblock

\vfill
\Title{Top-quark pair production in association with a $Z$ boson in the 4$\ell$ channel with the ATLAS experiment}
\vfill
\Author{Sebastian Heer\support, on behalf of the ATLAS Collaboration\copyright}
\Address{\institute}
\vfill
\begin{Abstract}

	The cross section of the \ttZ~and \ttW~processes are measured in a
	simultaneous fit using 36.1 $\text{fb}^{-1}$ of of proton--proton collisions
	at a centre-of-mass energy of $\sqrt{s}=13\ \TeV$ recorded by the ATLAS
	experiment at the LHC. In addition, a fit is performed in the 4$\ell$ channel
	only, resulting in a cross section of $\sigma_{\ttZ} = 1.07 \pm 0.26
	\,\text{pb}$. This result is consistent with the combined fit and agrees with
	the prediction by the Standard Model.

\end{Abstract}
\vfill
\begin{Presented}
$11^\mathrm{th}$ International Workshop on Top Quark Physics\\
Bad Neuenahr, Germany, September 16--21, 2018
\end{Presented}
\vfill
\end{titlepage}
\def\thefootnote{\fnsymbol{footnote}}
\setcounter{footnote}{0}

\section{Introduction}

The \ttZ~processes provides direct access to the weak coupling of the top quark
to the $Z$ boson. It is an important background in searches involving final
states with multiple leptons and $b$-quarks. Previous measurements by the
ATLAS and CMS experiments~\cite{atlas,Chatrchyan:2008aa} indicate agreement with the Standard
Model~\cite{Aaboud:2016xve,Sirunyan:2017uzs}.

Typically the \ttZ~process is measured in a two-dimensional fit together with
the \ttW~process. One of the analysis channels targeting the \ttZ~process has
four isolated prompt leptons. This final state provides the best
signal-to-background ratio among all analysis channels. Due to the branching
ratios of the top quark and the $Z$ boson, the number of expected events is
small in this analysis channel.

The dataset of 36.1 $\text{fb}^{-1}$ collected in 2015
and 2016 by the ATLAS detector is considered for the presented analysis. More
details about the analysis can be found in Reference~\cite{confnote}. 

\section{Tetralepton analysis}

Separate analysis channels are defined to target the \ttZ~and \ttW~processes.
This document focuses on the $4\ell$ channel, which is sensitive to the
\ttZ~process. Both $W$ bosons, resulting from the top-quark decays, and the $Z$
boson are required to decay leptonically. The most dominant background
processes arise from the diboson production, the production of a single top
quark in association with a $W$ and a $Z$ boson as well as fake leptons. A fit
is performed in the 4$\ell$ channel to extract the signal strength as well as
the background normalization of the $ZZ$ process.

Events with two pairs of opposite-sign leptons are selected, and at least one
pair must have the same flavour. Among the lepton pairs with opposite charge
and same flavour (OSSF), the pair with reconstructed invariant mass closest to
$m_Z$ is attributed to the $Z$ boson decay and denoted in the following by
$Z_1$.  The two remaining leptons are used to define $Z_2$.  Four signal
regions are defined according to the relative flavour of the two $Z_2$ leptons,
different flavour (DF) or same flavour (SF), and the number of $b$-tagged jets:
one, or at least two ($1b$, $2b$). The four signal regions are denoted as
4$\ell$-SF-1b, 4$\ell$-SF-2b, 4$\ell$-DF-1b and 4$\ell$-DF-2b. 

In the same-flavour regions, requirements on $E_\text{T}^{\text{miss}}$ are
applied in order to suppress the $ZZ$ background. This requirement depends on whether
the invariant mass of the $Z_2$ is close to the mass of the $Z$ boson. To
suppress events with fake leptons in the 1 $b$-tag multiplicity regions,
additional requirements on the scalar sum of the transverse momenta of the
third and fourth leptons ($p_{\text{T}34}$) are imposed. In other regions a
requirement on the transverse momentum of all leptons needs to be satisfied,
instead. The definitions of all signal regions are given in
Table~\ref{tab:4lsigregions}.

\begin{table}[htbp]
\centering \renewcommand{\arraystretch}{1.2}
	\caption{\label{tab:4lsigregions} Definitions of the four signal regions in the 4$\ell$ channel targeting the \ttZ~process~\cite{confnote}.
\vspace{1ex}}
\resizebox{\columnwidth}{!}{
\begin{tabular}{lcccr@{}cc@{}lc}
\toprule
Region & $Z_2$ leptons & $p_{\text{T}4}$ & $p_{\text{T}34}$ && $|m_{Z_{2}} - m_Z| $ & $E_\text{T}^{\text{miss}}$ && $N_{b{\text{-tagged jets}}}$\\
\midrule
	4$\ell$-DF-1b & $e^{\pm}\mu^{\mp}$ & - & $>$ \text{35 \GeV} &&-&-&& 1\\
	4$\ell$-DF-2b & $e^{\pm}\mu^{\mp}$ & $>$ \text{10 \GeV}& -&&-&-&& $\ge2$\\
	4$\ell$-SF-1b & $e^{\pm}e^{\mp},\mu^{\pm}\mu^{\mp}$ & - & $> \text{25 \GeV}$ & \begin{tabular}{@{}r@{}} \ldelim\{{2}{2ex} \\ \\ \end{tabular} &
		\begin{tabular}{c}$> \text{10 \GeV}$\\ $< \text{10 \GeV} $\end{tabular} & \begin{tabular}{c}$> \text{40 \GeV}$\\ $>\text{80 \GeV}$\end{tabular} &\begin{tabular}{@{}l@{}} \rdelim\}{2}{2ex} \\ \\ \end{tabular} & 1\\
			4$\ell$-SF-2b & $e^{\pm}e^{\mp},\mu^{\pm}\mu^{\mp}$ & $> \text{10 \GeV}$ & -&\begin{tabular}{@{}r@{}} \ldelim\{{2}{2ex} \\ \\ \end{tabular} &
				\begin{tabular}{c}$> \text{10 \GeV}$\\ $<\text{10 \GeV}$\end{tabular} & \begin{tabular}{c}-\\ $> \text{40 \GeV}$\end{tabular} &\begin{tabular}{@{}l@{}} \rdelim\}{2}{2ex} \\ \\ \end{tabular} & $\ge2$ \\
\bottomrule
\end{tabular}
}
\end{table}


A control region used to determine the $ZZ$ normalization, referred to as
$4\ell$-ZZ-CR is defined to have exactly four reconstructed leptons, a $Z_2$
pair with OSSF leptons, the value of both $m_{Z_1}$ and $m_{Z_2}$ within
$10\,\GeV$ of the mass of the $Z$ boson, and $20\,\GeV<
E_\text{T}^{\text{miss}} <40\,\GeV$. The normalization of the $ZZ$ background
is a free parameter in the fit. 

\begin{figure}[htbp]
\centering
  \subfloat[signal regions combined]{\includegraphics[width=0.5\textwidth]{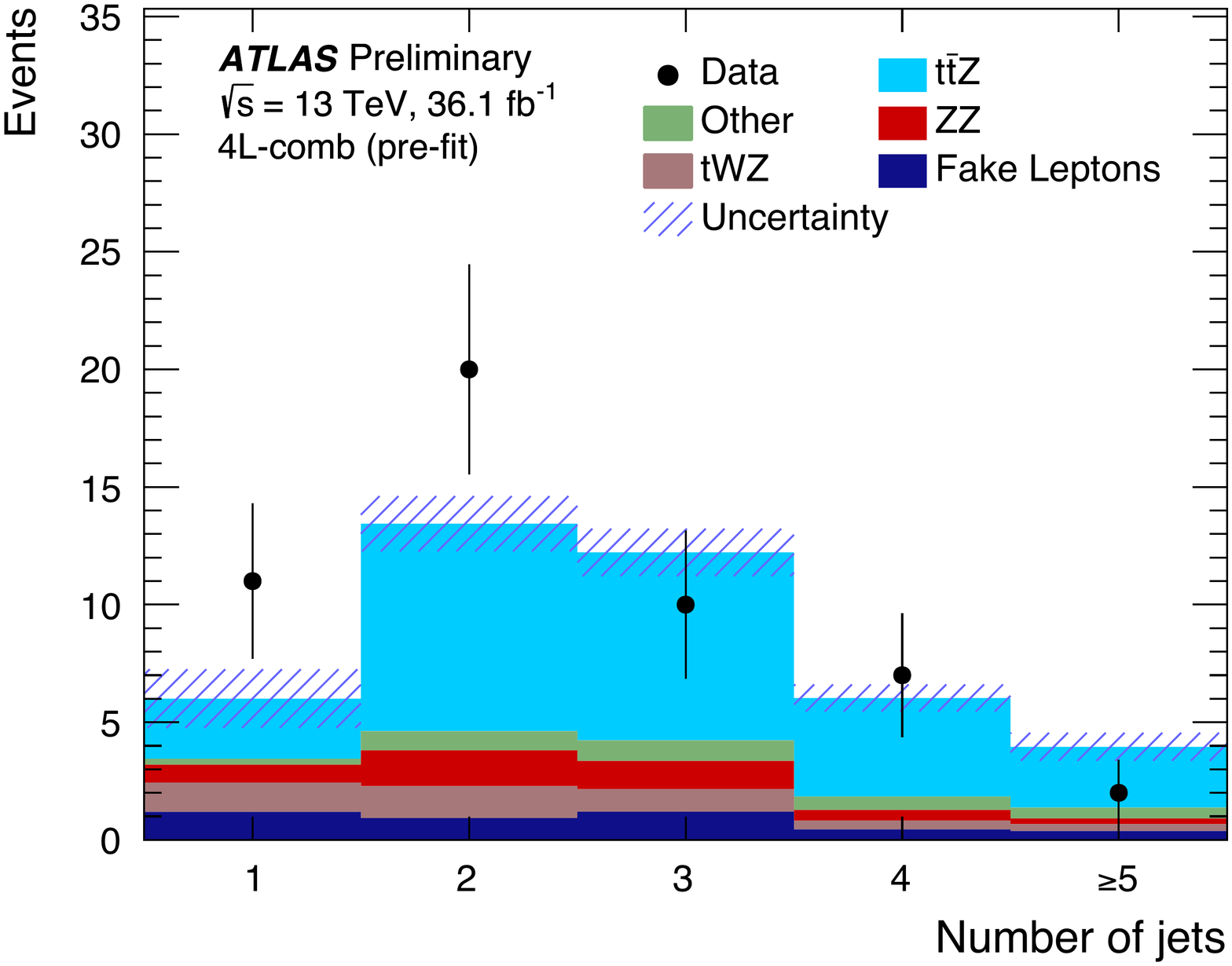}}
  \subfloat[$ZZ$ control region]{\includegraphics[width=0.5\textwidth]{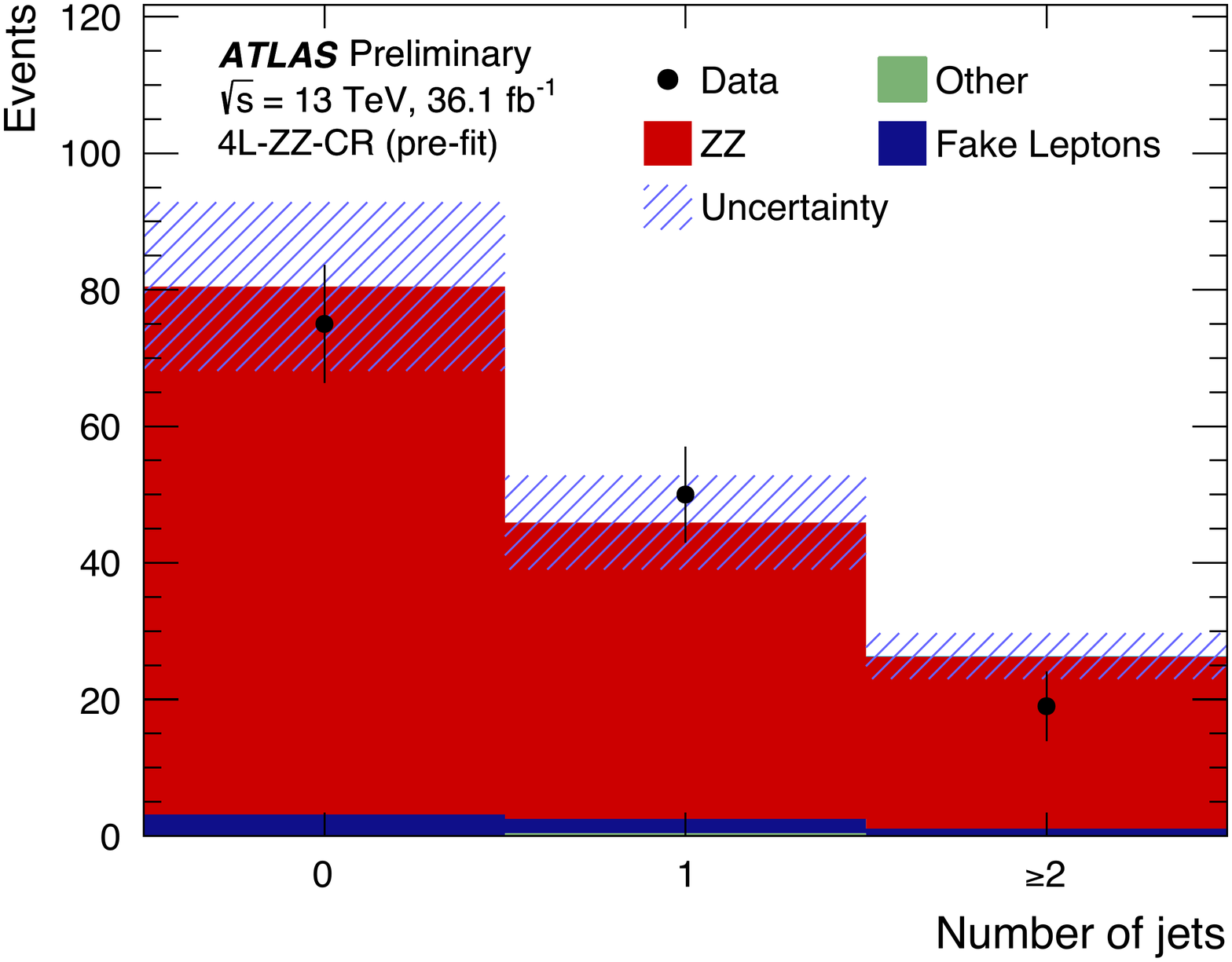}}
	\caption{Distribution of the number of jets in the signal regions (left) and control region (right). The data distributions are compared to MC predictions before the fit~\cite{confnote}.}
  \label{fig:nJets}
\end{figure}

Figure~\ref{fig:nJets} shows the data compared to the expected distributions
for all four signal regions combined, as well as for the control region,
showing good agreement between data and expectation.

The contribution from backgrounds containing fake leptons is estimated from
simulation and corrected with scale factors determined in two control regions:
one region enriched in $t\bar{t}$ events and one region enriched in $Z$+jets
events.  The scale factors are calibrated separately for electron and muon
fake-lepton candidates, and separately for leptons arising from heavy flavour
hadrons and other sources. Therefore, four scale factors are determined in
total.  The scale factors are applied to all simulated events with fewer
than four prompt leptons and depend on the number, flavour and origin of the
fake leptons.  It is verified that the scale factors for different generators
used in the simulation are consistent with each other.

\section{Results}

A binned maximum-likelihood fit is performed, which includes the signal regions
and the control region of the 4$\ell$ channel to extract the signal strength

\begin{equation}
	\mu_{\ttZ} = \frac{\sigma_{\ttZ}}{\sigma_{\ttZ}^{\text{SM}}}.
\end{equation}

\begin{figure}[htbp]
\centering
\includegraphics[width=0.85\textwidth]{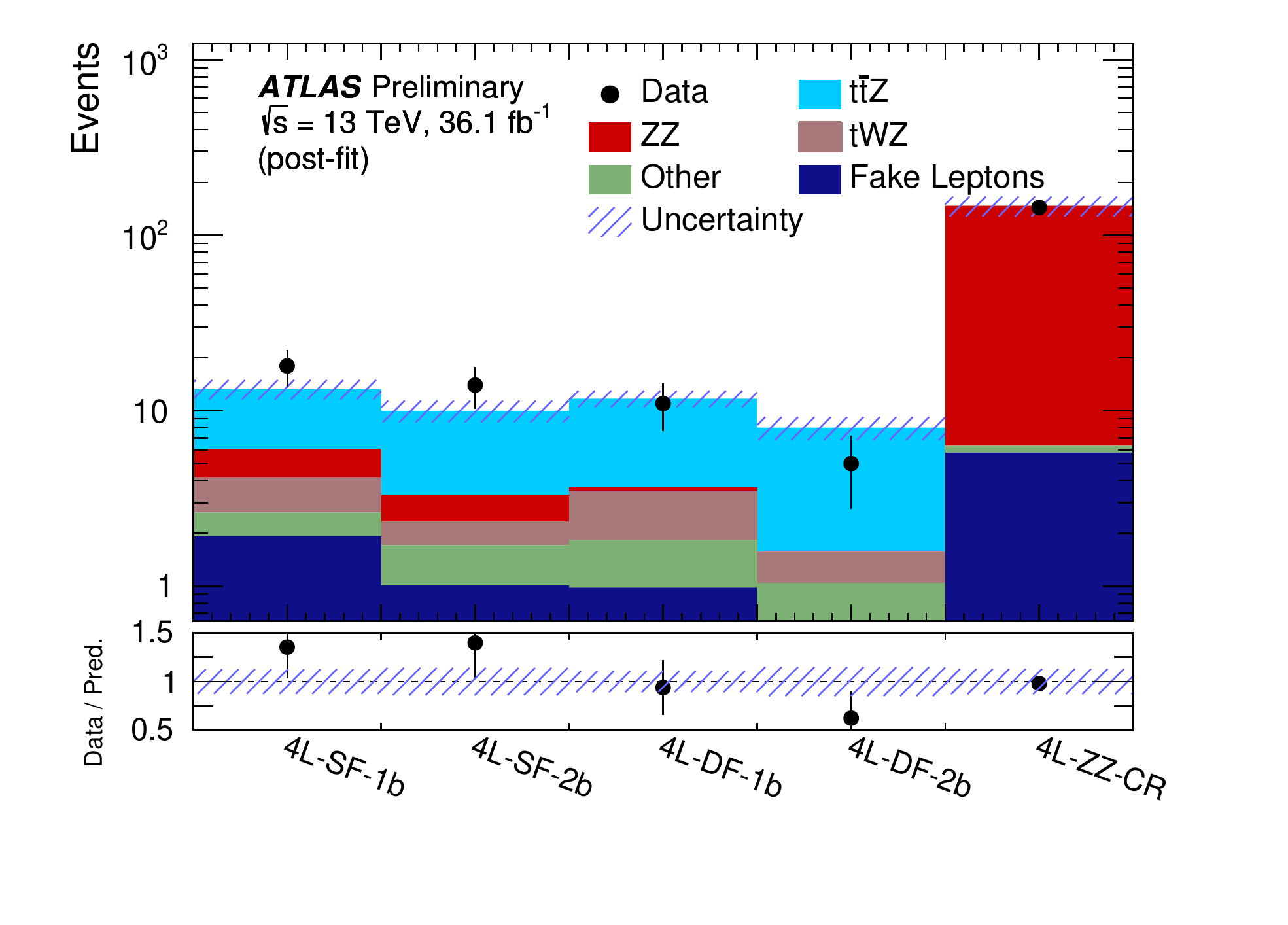}
	\caption{Data compared to the fit that extracts the \ttZ~cross section in the tetralepton regions after the fit has been performed. The background denoted as ``Other'' summarizes all minor Standard Model backgrounds with four reconstructed leptons~\cite{confnote}.}
\label{fig:summary}
\end{figure}

In all regions, good agreement between observed values and the expectation is
observed. Figure~\ref{fig:summary} shows all regions included in the fit after
the fit has been performed.

The normalization correction for the $ZZ$ background with respect to
the prediction is obtained from the fit and found to be compatible with unity:
$\mu_{ZZ} = 0.94 \pm 0.18$. The obtained signal strength of the \ttZ~process is

\begin{equation}
  \label{eq:mu4l}
\mu_{\ttZ} = 1.21 \pm 0.29
\end{equation}

and both the observed and expected significances are found to be larger than 5 standard deviations. Using
the theoretical prediction of $\sigma_{\ttZ} = 0.88^{+0.09}_{-0.11}$
pb~\cite{Alwall:2014hca}, the extracted cross section from the fit in the
4$\ell$ channel is $\sigma_{\ttZ}^{4\ell} = 1.07 \pm 0.26,$ pb demonstrating good agreement between the measured
and predicted cross sections. The dominant systematic uncertainties are related
to the modeling of the signal and flavour tagging. 

\Acknowledgements
The work of the author is currently funded by the European Research Council
under the European Union’s Seventh Framework Programme ERC Grant Agreement n.
617185.

\FloatBarrier

\end{document}